\documentclass[twocolumn,showpacs,preprintnumbers,amsmath,amssymb]{revtex4}
\usepackage{citesort}
\usepackage{epsfig}
\usepackage{graphicx}
\usepackage{dcolumn}
\usepackage{bm}
\usepackage{amsmath}
\pdfoutput=1 

\newcommand{\be}{\begin{equation}}
\newcommand{\ee}{\end{equation}}
\renewcommand{\vec}[1]{{\bf #1}}

\newcommand{\sub}[1]{_{\mbox{\scriptsize #1}}}

\begin{document}

\preprint{APS/123-QED}

\title{
Estimation of phase diffusion rates in a condensate interferometer using
the Gross-Pitaevskii Equation
}
\author{A. Fallon, R. H. Leonard and C. A. Sackett}%
\affiliation{Physics Department, University of Virginia,
Charlottesville, VA 22904}
\email{sackett@virginia.edu}
\date{\today}

\begin{abstract}
Atom interferometers using Bose-Einstein condensates are fundamentally
limited by a phase diffusion process that arises from atomic
interactions. 
The Gross-Pitaevskii equation is here used to accurately calculate the 
diffusion rate for a Bragg interferometer.
It is seen to agree with a Thomas-Fermi approximation
at large atom numbers and a perturbative approximation at low
atom numbers. The diffusion times obtained are generally longer
than the coherence times observed in experiments to date.
\end{abstract}


\maketitle

\section{Introduction}

Atom interferometry is a useful technique for a variety of precision
measurements \cite{Berman97,Cronin09}. 
The extension of atom interferometry to Bose-Einstein
condensates offers several potential advantages, most particularly
the ability to create an interferometer using trapped atoms
\cite{Wang05,Garcia06,Horikoshi07,Jo07}. This
can permit long interaction times while avoiding the complication
of having the atoms fall in gravity. In addition, the trapping
potential can shape the atomic trajectories into rings or other
geometries so as to optimize the interferometer performance for a particular
purpose \cite{Arnold02}.

Different approaches to condensate interferometry can be considered,
based on the method used to split the initial condensate into two
spatially distinct wave packets. One successful method is to 
expose the condensate to an off-resonant standing wave laser beam with
wave number $k$ \cite{Wang05,Garcia06,Horikoshi07}.
Bragg scattering can then generate
momentum kicks of $\pm 2\hbar k$, producing two packets moving in 
opposite directions. The light pulse must be short and
non-adiabatic with respect to the atomic motion. 
This provides a simple and robust
splitting scheme that is similar to techniques often used with 
thermal atoms \cite{Berman97,Cronin09}.

One drawback of condensate interferometry is that atoms in a condensate
interact relatively strongly due to their high density compared to most
thermal atom sources. These interactions can lead to a process
known as phase diffusion 
\cite{Holland96,Lewenstein96,Wright96,Wong96,Castin97,Javanainen97,Leggett98}.
When a condensate with $N$ atoms is optically 
split, the resulting packets will have a mean atom number of $N/2$ but
Poissonian fluctuations $\Delta N \approx N^{1/2}$. 
Since the energy of an interacting packet depends on its atom number, 
the number fluctuations make the energy
uncertain. This leads in turn to uncertainty in the quantum phase
evolution and thus noise in the atom interferometer measurement.
The interaction phase increases with time, so the overall effect is
to impose a limit on the usable measurement time of a condensate interferometer.
As the ability to achieve long interaction times is one of the key
advantages of a condensate interferometer, it is imperative to know
how stringent the phase diffusion limit really is.

Progress on this question has recently been made by
Ilo-Okeke and Zozulya (IZ) \cite{IloOkeke10}. 
They establish that in the limit of large atom
number $N$, the interferometer visibility $V$ can be well approximated
by
\be
\label{IZvis}
V = \exp(-2\xi^2 N)
\ee
where $\xi$ is an interaction phase
\be
\label{xidef}
\xi = \frac{1}{\hbar}\int_0^T g\,dt
\ee
and $g$ parametrizes the interactions as
\be
\label{gdef}
g = \frac{2\pi\hbar^2a}{m}\int |\psi|^4\,d^3r.
\ee
Here $T$ is the interferometer measurement time,
$a$ is the s-wave scattering length, $m$ is the atomic mass,
and $\psi(\vec{r},t)$ is the wave function of a packet in the interferometer,
normalized to one.

To get an actual value for the diffusion rate, IZ used a 
simple Thomas-Fermi approximation for $\psi$ in \eqref{gdef}. 
While this should predict how the visibility scales
with various parameters, the actual decoherence time 
might differ by a significant factor from this estimate.
In this paper, we present higher accuracy predictions for the
phase diffusion rate by numerically solving the Gross-Pitaevskii equation
to obtain a realistic $\psi(\vec{r},t)$. We are able to confirm some of the 
assumptions in the IZ derivation, and we can compare the more precise results
to the simpler Thomas-Fermi model. We find that the Thomas-Fermi result
is accurate in the limit of strong interactions,
but it naturally becomes less useful as the interaction strength is 
reduced. In the weakly interacting limit, a perturbative result
becomes accurate instead.
Most experiments to date fall in the crossover region between these
regimes.

\section{Gross-Pitaevskii Equation}

In the mean field approximation, the packet wave functions
are governed by the Gross-Pitaevskii equation \cite{Dalfovo99}
\be
\label{gpe}
i\hbar\frac{d\psi}{dt} = \left[
-\frac{\hbar^2}{2m}\nabla^2 + V\sub{trap}(\vec{r}) + \frac{4\pi\hbar^2a}{m} 
|\psi|^2\right]\psi.
\ee
Here $V\sub{trap}(\vec{r})$ represents the three dimensional trapping potential
in which the packets move. 
In accord with most experimental efforts, 
we assume it to be harmonic,
\be
V\sub{trap}(\vec{r}) = \frac{m}{2}\left(\omega_x^2 x^2 
+\omega_y^2 y^2 + \omega_z z^2\right).
\ee
The mean trap frequency is then 
$\omega \equiv (\omega_x\omega_y\omega_z)^{1/3}$.
We consider $^{87}$Rb atoms with scattering length $a = 5.77$~nm.

In an interferometer experiment, the initial condensate is the equilibrium
solution to \eqref{gpe} with $N$ atoms, $\psi_0(\vec{r})$. The optical
pulse splits this into two packets $\psi_+$ and $\psi_-$ moving at velocities
$\pm 2\hbar k/m$ along $z$. As the packets move apart, they initially
interact with each other. However, in a long-duration interferometer
the time that the packets interact is only a small fraction of the total.
We therefore neglect this interaction effect, as did IZ.
Instead we solve \eqref{gpe} for a single packet $\psi_+(\vec{r},t)$
with an atom number $N_+$ typically near $N/2$, and an initial state 
$\psi_+(\vec{r},0) = \psi_0(\vec{r})$. 
By using a Galilean transformation to the frame of the moving packet, we
can treat $\psi_+$ as being at rest.
However, since the equilibrium wave function
depends on the atom number, the time evolution of $\psi_+$ remains nontrivial.

To perform the calculation, we first determine $\psi_0$ using an
imaginary-time propagation technique \cite{Chiofalo00}. We use the Thomas-Fermi
wave function as an initial guess and propagate until the energy
converges. The atom number is then suddenly
reduced to $N_+$ and the equation is propagated forward in real time.

For propagation, 
we use the Strang-splitting technique outlined in \cite{Bao03}, which is
a mixed Fourier and real-space approach. We use a three-dimensional
grid of $128^3$ points, restricted to one octant by symmetry. Depending
on the parameters, a simulation takes about an hour to run on a 
desktop computer.

\begin{figure}
\includegraphics[width=3in]{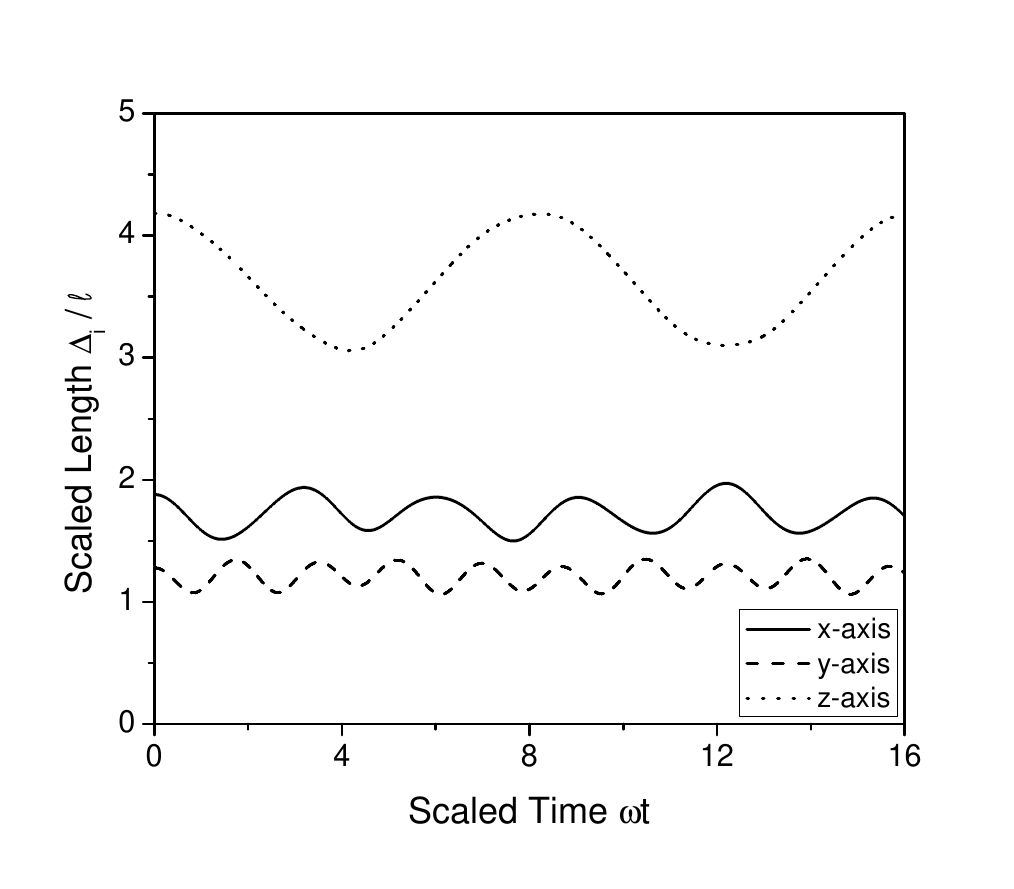}
\caption{\label{packet}
Packet dynamics in a condensate interferometer. 
The graph shows the results obtained by numerical solution of
the Gross-Pitaevskii equation for a $^{87}$Rb condensate with initial atom 
number $N = 10^4$ in a harmonic trap with frequencies
$(\omega_x, \omega_y, \omega_z) = 2\pi\times(3, 5, 1.2)$~Hz. Here
the $\Delta_i$ are the 
root-mean-square packet widths along direction $i$, scaled by
the harmonic oscillator length $\ell = \sqrt{\hbar/m\omega}$. At time $t=0$ the
number of atoms is suddenly reduced by a factor of two.
}
\end{figure}

Figure~\ref{packet} shows a typical result for the rms packet width
as a function of time. Since the packet is initially out of equilibrium,
we observe large amplitude oscillations in all three dimensions. The
dominant observed frequencies are the quadrupole modes \cite{Dalfovo99},
but these oscillations are anharmonic, leading to the complex structure
shown.

\section{Phase Diffusion}

The phase diffusion effect does not appear in the Gross-Pitaevskii equation
\eqref{gpe}, since phase diffusion requires a superposition of different
atom numbers. However, the
solution for the packet wave function $\psi_+$ makes evaluation of
the IZ visibility, Eq.~\eqref{IZvis}, straightforward. In fact, the IZ
calculation assumes that the wave functions $\psi_\pm$ 
do not depend on the precise value of $N_\pm$, 
and that the number sensitivity enters only through the phase evolution.
This assumption is essential for obtaining the simple result
\eqref{IZvis}.
On the one hand, it is reasonable since the
fluctuations in atom number, of order $\sqrt{N}$, are very small compared
to $N_+$ when $N\gg 1$. On the other hand, the Gross-Pitaevskii
equation does depend on the atom number, so after a sufficiently long time 
the wave functions $\psi_+$ and $\psi_-$
should begin to diverge if $N_+ \neq N_-$. 
These views are consistent if the time required for the wave functions
to diverge is long compared to the phase diffusion time. 

\begin{figure}
\includegraphics[width=3in]{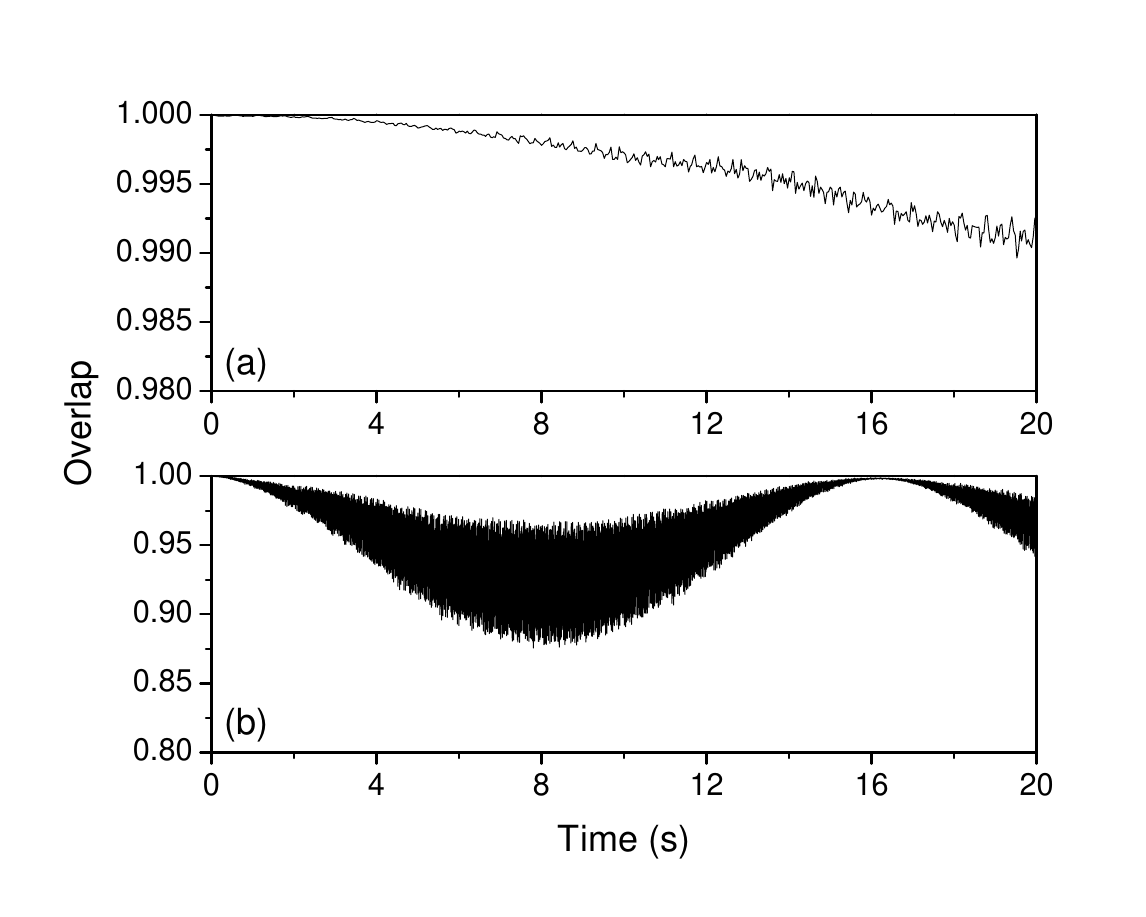}
\caption{\label{overlap}
Dependence of wave packet evolution on condensate number.
The plots show the overlap $|\langle\psi_+|\psi_-\rangle|^2$
between two wave packets evolving as in Fig.~\protect\ref{packet},
but where $\psi_\pm$ has an atom number of $N/2 \pm \sqrt{N}$.
Graph (a) uses $N = 10^4$ $^{87}$Rb atoms and
$(\omega_x,\omega_y,\omega_z) = 2\pi\times (3, 5, 0.2)$~Hz, corresponding
to the experiments of \protect\cite{Burke10}. Graph (b) uses
$N = 3\times 10^3$ $^{87}$Rb atoms and
$(\omega_x,\omega_y,\omega_z) = 2\pi\times (60, 60, 17)$~Hz, corresponding
the experiments of \protect\cite{Horikoshi07}. 
The curve in (b) appears thick due to rapid oscillations 
at the trap frequency time scale.}
\end{figure}

We were able to verify this assumption,
by calculating
the overlap $|\langle \psi_+|\psi_-\rangle|^2$
between a packet
with $N_+ = N/2 + \sqrt{N}$ and another with
$N_- = N/2 - \sqrt{N}$. The results are shown in Fig.~\ref{overlap}
for two sets of parameters. The first 
corresponds to the experiments of \cite{Garcia06}.
Here the overlap remains close to one at all plausible times. The second
parameters correspond to the experiments
of \cite{Horikoshi07}. Here the overlap does decrease, though only by about
15\% at most. The time scale for the decrease is several seconds,
which is about an order of magnitude longer than the diffusion time
calculated below. The revival of the overlap at $T = 17$~s is due to 
a rephasing of the packet oscillation modes.

These two sets of parameters represent 
the weakest and strongest interaction strengths so far realized
in this type of experiment. We therefore confirm that under typical
conditions, the assumption used by IZ is valid.

IZ proposed a Thomas-Fermi approximation for $\xi$ in \eqref{xidef},
by taking $\psi_+(\vec{r},t)$ to be the stationary Thomas-Fermi
wave function for $N_+ = N/2$ atoms. This leads to
\be
\xi \approx 0.64 \left(\frac{a}{\ell}\right)^{2/5}\omega T N^{-3/5}
\ee
where $\ell \equiv (\hbar/m\omega)^{1/2}$
and the numerical factor is $(1800/16807)^{1/5}$.
It is also possible to evaluate $\xi$ in the limit of small
interactions, where $\psi_+$ can be approximated by the non-interacting
harmonic oscillator wave function. Here we obtain
\be
\xi \approx \frac{1}{\sqrt{2\pi}} \frac{a}{\ell}\omega T.
\ee

In both of these cases, $\xi$ depends linearly on $T$, 
leading to a visibility that decays as
\be
V = \exp\left(-\frac{2T^2}{\tau^2}\right)
\ee
for a diffusion time $\tau$.
The Thomas-Fermi approximation gives 
\be
\label{tauTF}
\tau\sub{TF} = 1.56\left(\frac{\ell}{a}\right)^{2/5} \frac{N^{1/10}}{\omega}.
\ee
while the weak interaction limit gives
\be
\label{tau0}
\tau_0 = \sqrt{2\pi}\frac{\ell}{a\omega N^{1/2}}.
\ee

\begin{figure}
\includegraphics[width=3in]{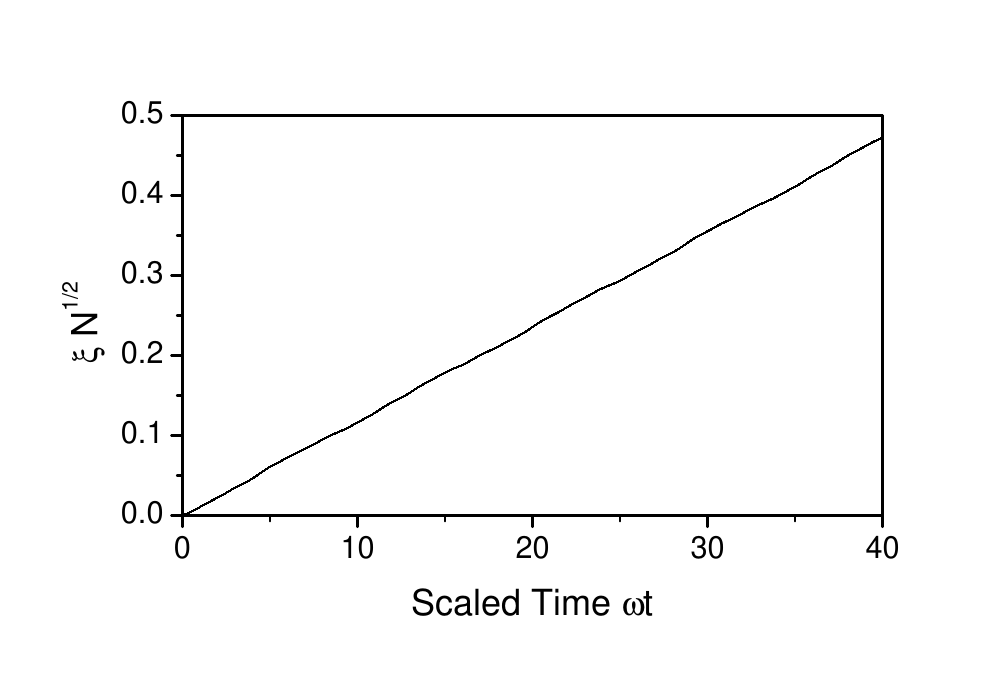}
\caption{\label{xi}
Time evolution of the interaction phase. 
The phase parameter $\xi$ of \protect\eqref{xi} is calculated for
$N = 10^4$~$^{87}$Rb atoms in a trap with
$(\omega_x,\omega_y,\omega_z) = 2\pi\times (3, 5, 1.2)$~Hz. 
The dominant behavior is the linear increase in time. The oscillations
around this behavior 
are small enough to neglect without significant loss of accuracy.}
\end{figure}

In general, however, $\xi$ has a more complicated time dependence
since $g$ can exhibit oscillations as the packet evolves.
These oscillations are in fact fairly modest under all the conditions
we considered, as illustrated in Fig.~\ref{xi}.
We therefore approximate $\xi$ with a linear fit and obtain a diffusion time 
\be
\label{tau}
\tau= N^{-1/2}\left<\frac{d\xi}{dt}\right>^{-1}.
\ee
The brackets here represent a time average as determined by the slope
of the fit line.

\section{Diffusion Time Results}

\begin{figure}
\includegraphics[width=3in]{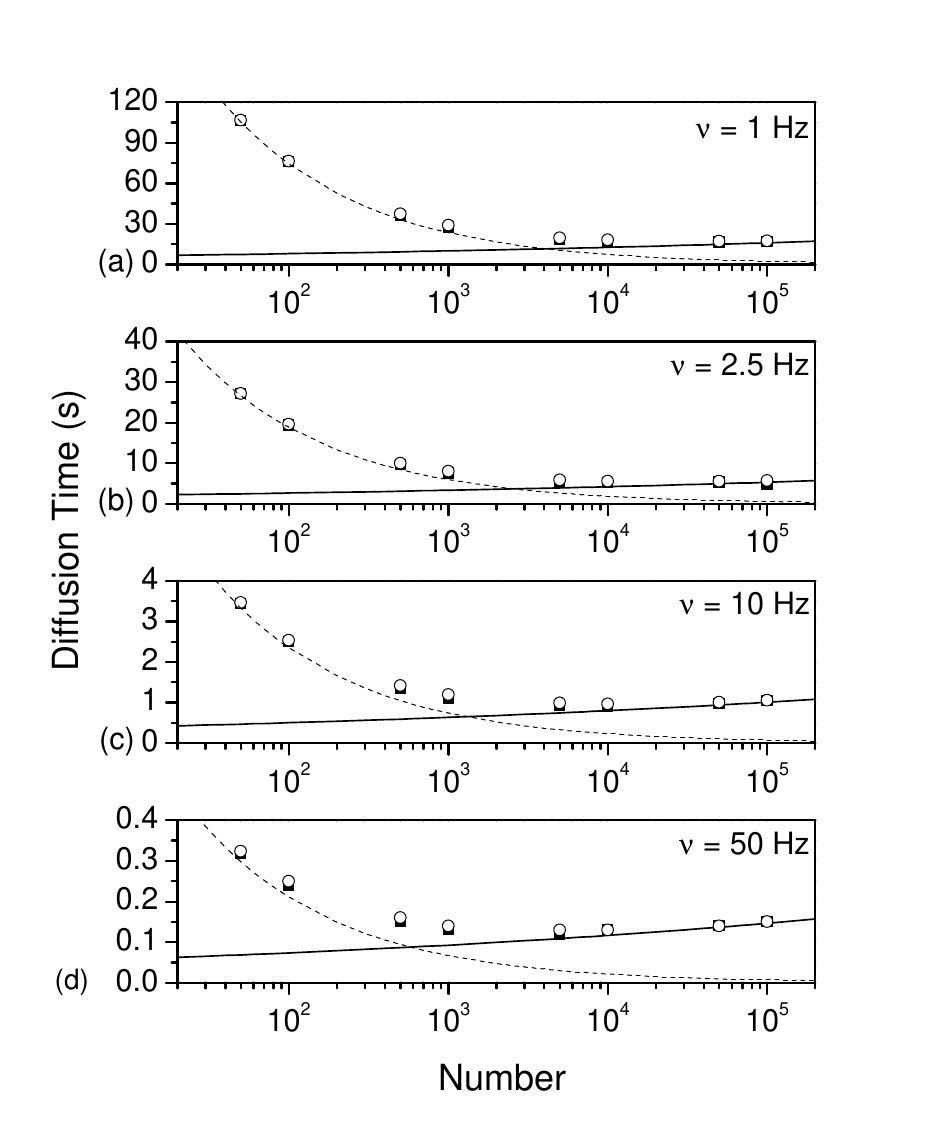}
\caption{\label{diffusion_times}
Numerical results for phase diffusion time under various conditions
indicated. Here $\nu$ indicates the mean trap frequency $\omega/2\pi$.
Open circles use $\omega_x = \omega_y = \lambda \omega_z$ for
$\lambda = 5$,
and closed squares use $\lambda = 1$. The solid curves show the
approximate Thomas-Fermi result and the dashed curves show the weakly
interacting approximation.}
\end{figure}

We investigated how the diffusion time depends on the various parameters.
Figure~\ref{diffusion_times} illustrates
the dependence on atom
number $N$, trap frequency $\omega$, 
and trap asymmetry $\lambda = \omega_x/\omega_z$ for a cylindrical
trap with $\omega_y = \omega_x$. The general trends agree with expectations:
The diffusion time scales rapidly with $\omega$ since a tighter trap
leads to higher atom density.
For large atom numbers, $\tau$ increases slowly with $N$ owing to the
interplay between the increasing density in the Thomas-Fermi wave
function and the decreasing relative 
impact of the $\sqrt{N}$ number fluctuations.
The diffusion time also increases at low atom numbers as the interaction
are reduced. The diffusion time depends only weakly on the trap symmetry,
in accord with the Thomas-Fermi result and weak-interaction results.

\begin{figure}
\includegraphics[width=3in]{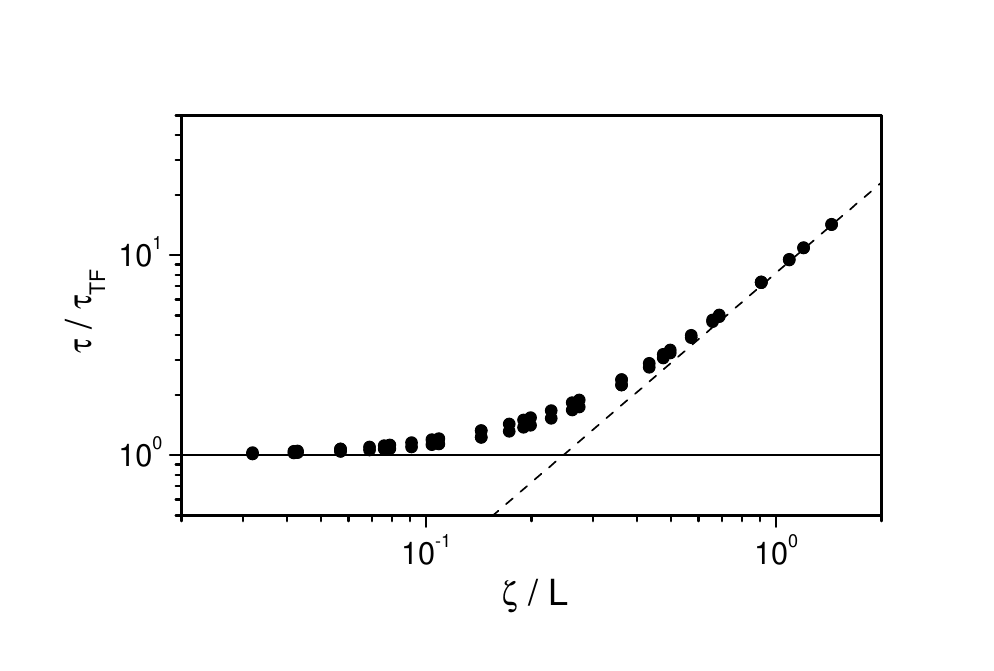}
\caption{\label{TFcomp}
Relation between phase diffusion time and the Thomas-Fermi
parameter $\zeta/L$, where $\zeta$ is the healing length
and $L$ is the Thomas-Fermi radius. The data
are scaled by the Thomas-Fermi diffusion time from 
\protect\eqref{tauTF}, and the solid line shows $\tau = \tau_{TF}$.
The dashed line shows the weak-interaction approximation of
\protect\eqref{tau0}.
Data points show calculated results for a mix of asymmetries $\lambda = 1$ and
$\lambda = 5$.
}
\end{figure}

As seen, the results agree well with the
Thomas-Fermi approximation at larger $N$, where the approximation
is expected to hold. At low $N$, the results converge to the
weak-interaction result. The crossover between these
regimes is governed by the healing length 
$\zeta = (8\pi na)^{-1/2}$, for the atomic density $n$. 
The Thomas-Fermi approximation is
good when $\zeta$ is small compared to the 
wavepacket size, and interactions are weak when $\zeta$ is large.
To illustrate this, Fig.~\ref{TFcomp}
plots the ratio $\tau/\tau\sub{TF}$ as a function of $\zeta/L$, where
$L = (L_x L_y L_z)^{1/3}$ is the mean condensate size.
Here we used the Thomas-Fermi approximations for $\zeta$
and $L$, via \cite{Dalfovo99}
\be
\label{TFL}
L  = (15N a\ell^4)^{1/5}
\ee
and the peak density
\be
n = \frac{1}{8\pi} \frac{L^2}{a\ell^4},
\ee
from which $\zeta/L= (\ell/15Na)^{2/5}$. 
As seen, the diffusion time can be rather accurately represented as a function
of the single variable $\zeta/L$.

\begin{figure}
\includegraphics[width=3in]{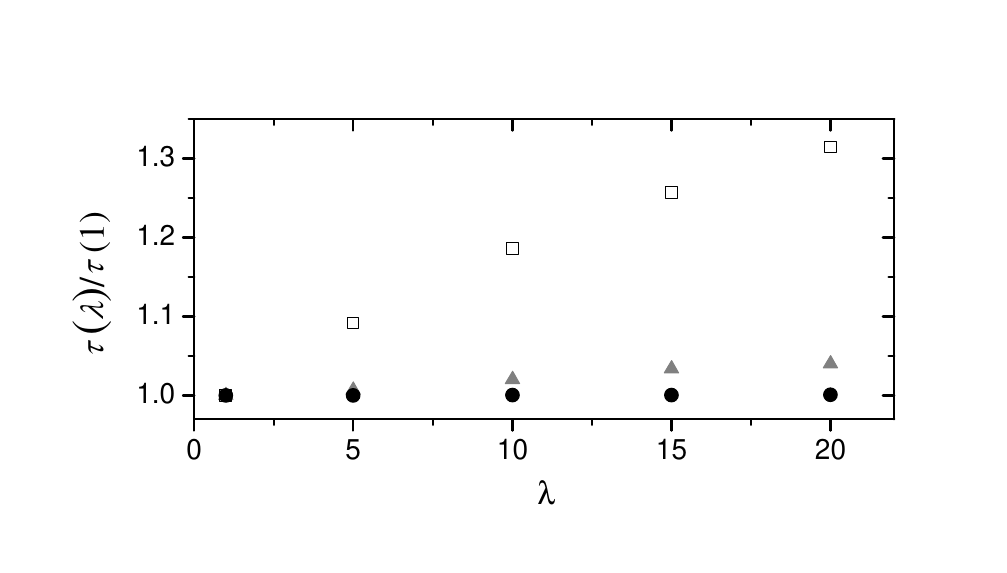}
\caption{\label{lambda}
Dependence of diffusion time $\tau(\lambda)$
on trap symmetry $\lambda = \omega_x/\omega_z$
for a cylindrical trap with $\omega_x = \omega_y$. Values are scaled
by the diffusion time for a symmetric trap, $\tau(1)$. Different
symbols are calculated for different interaction strengths, as characterized
by the ratio of the healing length $\xi$ to the trap size $\ell$.
Circles have $\xi \approx 5\ell$, squares have $\xi \approx 0.2\ell$, and
triangles have $\xi \approx 0.03\ell$.
}
\end{figure}

In both the strong and weak interaction limit, the diffusion time
is independent of the trap symmetry. A weak dependence is observed,
however, in the crossover region. This is illustrated in Fig.~\ref{lambda}.

\section{Impact on Experiments}

Table I shows the parameters used and coherence times obtained
in several experimental implementations, including the most recent
results from our own group. As seen from the $\zeta/L$ values,
all of these experiments are in the crossover regime where the
predicted coherence time is somewhat longer than either the Thomas-Fermi
or weak interaction formulas would predict. 

\begin{table*}
\begin{tabular}{llllllll}
Reference \hspace{2em} & $\nu_x$\,(Hz) \  & $\nu_y$\,(Hz) \  & $\nu_z$\,(Hz) \  
& $N$ \hspace{3em} & $T$\,(s) \hspace{1em} &  $\tau$\,(s) \hspace{1em} & \ $\zeta/L$  \\
\hline
Wang \protect\cite{Wang05,Olshanii05} & 
100 & 100 & 5 & 3000 & 0.01 & 0.23 & 0.14\\
Garcia \protect\cite{Garcia06} &
3.3 & 6 & 1.2 & 10,000 & 0.04 & 4.6 & 0.14 \\
Horikoshi 	\protect\cite{Horikoshi07} &
60 & 60 & 0.017 & 3000 & 0.06 & 0.17 & 0.13\\
Horikoshi \protect\cite{Horikoshi07} &
60 & 60 & 0.010 & 3000 & 0.1* & 0.23 & 0.14\\
Burke \protect\cite{Burke08} &
3.3 & 6 & 1.1 & 30,000 & 0.9* & 4.8 & 0.09\\
Segal \protect\cite{Segal10} &
80 & 80 & 4.1 & 3000 & 0.24* & 0.30 & 0.14\\
Current	&
3 & 5 & 0.2 & 10,000 & 0.10 & 13.5 & 0.16\\
\end{tabular}
\caption{
Summary of experimental atom interferometer performance, compared
to calculated diffusion times. All results are for $^{87}$Rb atoms.
Here the $\nu_i$ are the trap frequencies,
$N$ the atom numbers, and $T$ the experimental measurement time.
The asterisks denote measurements exhibiting interference with
uncontrolled shot-to-shot phase fluctuations. Our calculated diffusion
times are $\tau$, and $\zeta/L$ parametrizes the relative strength of
interactions.
}

\end{table*}

A long phase diffusion time is of course beneficial, but as can
be seen from the table, experiments to date have mostly been instead limited
by technical noise sources. The work by Segal {\em et al.}
\cite{Segal10} is a notable exception. The measurement time
in that experiment is close to the predicted diffusion time, and
in fact exceeds the approximate Thomas-Fermi result of 0.19~s.
In that experiment, the interferometer did not exhibit a stable
output, but fluctuated from one run to the next. The IZ theory
predicts such behavior when the measurement time approaches the diffusion
time. However, the observed fluctuations could 
also be explained by technical sources \cite{Segal10}.

Our results do suggest that substantial
performance improvements are possible. For instance,
if a system similar to our current one were configured to give a loop
geometry with a 2-cm radius and a 10-s coherence time, a single shot-noise
limited measurement would have a Sagnac rotation sensitivity
of about $3\times 10^{-9}$~rad/s. 

\section{Conclusions}

We have elaborated on Ilo-Okeke and Zozulya's analysis of phase
diffusion in an optically-coupled condensate interferometer, by
using a realistic simulation of the Gross-Pitaevskii equation to 
calculate the dynamical evolution of the interferometer wave packet.
We find that the diffusion time is well approximated by a Thomas-Fermi
result in the strongly-interacting limit, and by a perturbative result
in the weakly-interacting limit.
We hope that these results will be of use in
designing and interpreting future experiments based on condensate 
interferometry.

We thank Alex Zozulya for helpful comments.
This material is based upon work supported by the National 
Science Foundation under Grant Number 0969916.


\begin{thebibliography}{22}
\expandafter\ifx\csname natexlab\endcsname\relax\def\natexlab#1{#1}\fi
\expandafter\ifx\csname bibnamefont\endcsname\relax
  \def\bibnamefont#1{#1}\fi
\expandafter\ifx\csname bibfnamefont\endcsname\relax
  \def\bibfnamefont#1{#1}\fi
\expandafter\ifx\csname citenamefont\endcsname\relax
  \def\citenamefont#1{#1}\fi
\expandafter\ifx\csname url\endcsname\relax
  \def\url#1{\texttt{#1}}\fi
\expandafter\ifx\csname urlprefix\endcsname\relax\def\urlprefix{URL }\fi
\providecommand{\bibinfo}[2]{#2}
\providecommand{\eprint}[2][]{\url{#2}}

\bibitem[{\citenamefont{Berman}(1997)}]{Berman97}
\bibinfo{editor}{\bibfnamefont{P.~R.} \bibnamefont{Berman}}, ed.,
  \emph{\bibinfo{title}{Atom Interferometry}} (\bibinfo{publisher}{Academic
  Press}, \bibinfo{address}{San Diego}, \bibinfo{year}{1997}).

\bibitem[{\citenamefont{Cronin et~al.}(2009)\citenamefont{Cronin, Schmiedmayer,
  and Pritchard}}]{Cronin09}
\bibinfo{author}{\bibfnamefont{A.~D.} \bibnamefont{Cronin}},
  \bibinfo{author}{\bibfnamefont{J.}~\bibnamefont{Schmiedmayer}},
  \bibnamefont{and} \bibinfo{author}{\bibfnamefont{D.~E.}
  \bibnamefont{Pritchard}}, \bibinfo{journal}{Rev. Mod. Phys.}
  \textbf{\bibinfo{volume}{81}}, \bibinfo{pages}{1051} (\bibinfo{year}{2009}).

\bibitem[{\citenamefont{Wang et~al.}(2005)\citenamefont{Wang, Anderson, Bright,
  Cornell, Diot, Kishimoto, Prentiss, Saravanan, Segal, and Wu}}]{Wang05}
\bibinfo{author}{\bibfnamefont{Y.~J.} \bibnamefont{Wang}},
  \bibinfo{author}{\bibfnamefont{D.~Z.} \bibnamefont{Anderson}},
  \bibinfo{author}{\bibfnamefont{V.~M.} \bibnamefont{Bright}},
  \bibinfo{author}{\bibfnamefont{E.~A.} \bibnamefont{Cornell}},
  \bibinfo{author}{\bibfnamefont{Q.}~\bibnamefont{Diot}},
  \bibinfo{author}{\bibfnamefont{T.}~\bibnamefont{Kishimoto}},
  \bibinfo{author}{\bibfnamefont{M.}~\bibnamefont{Prentiss}},
  \bibinfo{author}{\bibfnamefont{R.~A.} \bibnamefont{Saravanan}},
  \bibinfo{author}{\bibfnamefont{S.~R.} \bibnamefont{Segal}}, \bibnamefont{and}
  \bibinfo{author}{\bibfnamefont{S.}~\bibnamefont{Wu}}, \bibinfo{journal}{Phys.
  Rev. Lett.} \textbf{\bibinfo{volume}{94}}, \bibinfo{pages}{090405}
  (\bibinfo{year}{2005}).

\bibitem[{\citenamefont{Garcia et~al.}(2006)\citenamefont{Garcia, Deissler,
  Hughes, Reeves, and Sackett}}]{Garcia06}
\bibinfo{author}{\bibfnamefont{O.}~\bibnamefont{Garcia}},
  \bibinfo{author}{\bibfnamefont{B.}~\bibnamefont{Deissler}},
  \bibinfo{author}{\bibfnamefont{K.~J.} \bibnamefont{Hughes}},
  \bibinfo{author}{\bibfnamefont{J.~M.} \bibnamefont{Reeves}},
  \bibnamefont{and} \bibinfo{author}{\bibfnamefont{C.~A.}
  \bibnamefont{Sackett}}, \bibinfo{journal}{Phys. Rev. A}
  \textbf{\bibinfo{volume}{74}}, \bibinfo{pages}{031601(R)}
  (\bibinfo{year}{2006}).

\bibitem[{\citenamefont{Horikoshi and Nakagawa}(2007)}]{Horikoshi07}
\bibinfo{author}{\bibfnamefont{M.}~\bibnamefont{Horikoshi}} \bibnamefont{and}
  \bibinfo{author}{\bibfnamefont{K.}~\bibnamefont{Nakagawa}},
  \bibinfo{journal}{Phys. Rev. Lett.} \textbf{\bibinfo{volume}{99}},
  \bibinfo{pages}{180401} (\bibinfo{year}{2007}).

\bibitem[{\citenamefont{Jo et~al.}(2007)\citenamefont{Jo, Shin, Will, Pasquini,
  Saba, Ketterle, Pritchard, Vengalattore, and Prentiss}}]{Jo07}
\bibinfo{author}{\bibfnamefont{G.-B.} \bibnamefont{Jo}},
  \bibinfo{author}{\bibfnamefont{Y.}~\bibnamefont{Shin}},
  \bibinfo{author}{\bibfnamefont{S.}~\bibnamefont{Will}},
  \bibinfo{author}{\bibfnamefont{T.~A.} \bibnamefont{Pasquini}},
  \bibinfo{author}{\bibfnamefont{M.}~\bibnamefont{Saba}},
  \bibinfo{author}{\bibfnamefont{W.}~\bibnamefont{Ketterle}},
  \bibinfo{author}{\bibfnamefont{D.~E.} \bibnamefont{Pritchard}},
  \bibinfo{author}{\bibfnamefont{M.}~\bibnamefont{Vengalattore}},
  \bibnamefont{and} \bibinfo{author}{\bibfnamefont{M.}~\bibnamefont{Prentiss}},
  \bibinfo{journal}{Phys. Rev. Lett.} \textbf{\bibinfo{volume}{98}},
  \bibinfo{pages}{030407} (\bibinfo{year}{2007}).

\bibitem[{\citenamefont{Arnold and Riis}(1999)}]{Arnold02}
\bibinfo{author}{\bibfnamefont{A.~S.} \bibnamefont{Arnold}} \bibnamefont{and}
  \bibinfo{author}{\bibfnamefont{E.}~\bibnamefont{Riis}}, \bibinfo{journal}{J.
  Mod. Optics} \textbf{\bibinfo{volume}{49}}, \bibinfo{pages}{5861}
  (\bibinfo{year}{1999}).

\bibitem[{\citenamefont{Holland et~al.}(1996)\citenamefont{Holland, Williams,
  Coakley, and Cooper}}]{Holland96}
\bibinfo{author}{\bibfnamefont{M.}~\bibnamefont{Holland}},
  \bibinfo{author}{\bibfnamefont{J.}~\bibnamefont{Williams}},
  \bibinfo{author}{\bibfnamefont{K.}~\bibnamefont{Coakley}}, \bibnamefont{and}
  \bibinfo{author}{\bibfnamefont{J.}~\bibnamefont{Cooper}},
  \bibinfo{journal}{Quantum and Semiclassical Optics}
  \textbf{\bibinfo{volume}{8}}, \bibinfo{pages}{571} (\bibinfo{year}{1996}).

\bibitem[{\citenamefont{Lewenstein and You}(1996)}]{Lewenstein96}
\bibinfo{author}{\bibfnamefont{M.}~\bibnamefont{Lewenstein}} \bibnamefont{and}
  \bibinfo{author}{\bibfnamefont{L.}~\bibnamefont{You}},
  \bibinfo{journal}{Phys. Rev. Lett.} \textbf{\bibinfo{volume}{77}},
  \bibinfo{pages}{3489} (\bibinfo{year}{1996}).

\bibitem[{\citenamefont{Wright et~al.}(1996)\citenamefont{Wright, Walls, and
  Garrison}}]{Wright96}
\bibinfo{author}{\bibfnamefont{E.~M.} \bibnamefont{Wright}},
  \bibinfo{author}{\bibfnamefont{D.~F.} \bibnamefont{Walls}}, \bibnamefont{and}
  \bibinfo{author}{\bibfnamefont{J.~C.} \bibnamefont{Garrison}},
  \bibinfo{journal}{Phys. Rev. Lett.} \textbf{\bibinfo{volume}{77}},
  \bibinfo{pages}{2158} (\bibinfo{year}{1996}).

\bibitem[{\citenamefont{Wong et~al.}(1996)\citenamefont{Wong, Collett, and
  Walls}}]{Wong96}
\bibinfo{author}{\bibfnamefont{T.}~\bibnamefont{Wong}},
  \bibinfo{author}{\bibfnamefont{M.~J.} \bibnamefont{Collett}},
  \bibnamefont{and} \bibinfo{author}{\bibfnamefont{D.~F.} \bibnamefont{Walls}},
  \bibinfo{journal}{Phys. Rev. A} \textbf{\bibinfo{volume}{54}},
  \bibinfo{pages}{R3718} (\bibinfo{year}{1996}).

\bibitem[{\citenamefont{Castin and Dalibard}(1997)}]{Castin97}
\bibinfo{author}{\bibfnamefont{Y.}~\bibnamefont{Castin}} \bibnamefont{and}
  \bibinfo{author}{\bibfnamefont{J.}~\bibnamefont{Dalibard}},
  \bibinfo{journal}{Phys. Rev. A} \textbf{\bibinfo{volume}{55}},
  \bibinfo{pages}{4330} (\bibinfo{year}{1997}).

\bibitem[{\citenamefont{Javanainen and Wilkens}(1997)}]{Javanainen97}
\bibinfo{author}{\bibfnamefont{J.}~\bibnamefont{Javanainen}} \bibnamefont{and}
  \bibinfo{author}{\bibfnamefont{M.}~\bibnamefont{Wilkens}},
  \bibinfo{journal}{Phys. Rev. Lett.} \textbf{\bibinfo{volume}{78}},
  \bibinfo{pages}{4675} (\bibinfo{year}{1997}).

\bibitem[{\citenamefont{Leggett and Sols}(1998)}]{Leggett98}
\bibinfo{author}{\bibfnamefont{A.~J.} \bibnamefont{Leggett}} \bibnamefont{and}
  \bibinfo{author}{\bibfnamefont{F.}~\bibnamefont{Sols}},
  \bibinfo{journal}{Phys. Rev. Lett.} \textbf{\bibinfo{volume}{81}},
  \bibinfo{pages}{1344} (\bibinfo{year}{1998}).

\bibitem[{\citenamefont{Ilo-Okeke and Zozulya}(2010)}]{IloOkeke10}
\bibinfo{author}{\bibfnamefont{E.~O.} \bibnamefont{Ilo-Okeke}}
  \bibnamefont{and} \bibinfo{author}{\bibfnamefont{A.~A.}
  \bibnamefont{Zozulya}}, \bibinfo{journal}{Phys. Rev. A}
  \textbf{\bibinfo{volume}{82}}, \bibinfo{pages}{053603}
  (\bibinfo{year}{2010}).

\bibitem[{\citenamefont{Dalfovo et~al.}(1999)\citenamefont{Dalfovo, Giorgini,
  Pitaevskii, and Stringari}}]{Dalfovo99}
\bibinfo{author}{\bibfnamefont{F.}~\bibnamefont{Dalfovo}},
  \bibinfo{author}{\bibfnamefont{S.}~\bibnamefont{Giorgini}},
  \bibinfo{author}{\bibfnamefont{L.}~\bibnamefont{Pitaevskii}},
  \bibnamefont{and}
  \bibinfo{author}{\bibfnamefont{S.}~\bibnamefont{Stringari}},
  \bibinfo{journal}{Rev. Mod. Phys.} \textbf{\bibinfo{volume}{71}},
  \bibinfo{pages}{463} (\bibinfo{year}{1999}).

\bibitem[{\citenamefont{Chiofalo et~al.}(2000)\citenamefont{Chiofalo, Succi,
  and Tosi}}]{Chiofalo00}
\bibinfo{author}{\bibfnamefont{M.~L.} \bibnamefont{Chiofalo}},
  \bibinfo{author}{\bibfnamefont{S.}~\bibnamefont{Succi}}, \bibnamefont{and}
  \bibinfo{author}{\bibfnamefont{M.~P.} \bibnamefont{Tosi}},
  \bibinfo{journal}{Phys. Rev. E} \textbf{\bibinfo{volume}{62}},
  \bibinfo{pages}{7438} (\bibinfo{year}{2000}).

\bibitem[{\citenamefont{Bao et~al.}(2003)\citenamefont{Bao, Jaksch, and
  Markowich}}]{Bao03}
\bibinfo{author}{\bibfnamefont{W.}~\bibnamefont{Bao}},
  \bibinfo{author}{\bibfnamefont{D.}~\bibnamefont{Jaksch}}, \bibnamefont{and}
  \bibinfo{author}{\bibfnamefont{P.}~\bibnamefont{Markowich}},
  \bibinfo{journal}{J. Comp. Phys.} \textbf{\bibinfo{volume}{187}},
  \bibinfo{pages}{318} (\bibinfo{year}{2003}).

\bibitem[{\citenamefont{Burke}(2010)}]{Burke10}
\bibinfo{author}{\bibfnamefont{J.~H.~T.} \bibnamefont{Burke}}, Ph.D. thesis,
  \bibinfo{school}{University of Virginia} (\bibinfo{year}{2010}).

\bibitem[{\citenamefont{Olshanii and Dunjko}(2005)}]{Olshanii05}
\bibinfo{author}{\bibfnamefont{M.}~\bibnamefont{Olshanii}} \bibnamefont{and}
  \bibinfo{author}{\bibfnamefont{V.}~\bibnamefont{Dunjko}}
  (\bibinfo{year}{2005}), \bibinfo{note}{eprint cond-mat/0505358}.

\bibitem[{\citenamefont{Burke et~al.}(2008)\citenamefont{Burke, Deissler,
  Hughes, and Sackett}}]{Burke08}
\bibinfo{author}{\bibfnamefont{J.~H.~T.} \bibnamefont{Burke}},
  \bibinfo{author}{\bibfnamefont{B.}~\bibnamefont{Deissler}},
  \bibinfo{author}{\bibfnamefont{K.~J.} \bibnamefont{Hughes}},
  \bibnamefont{and} \bibinfo{author}{\bibfnamefont{C.~A.}
  \bibnamefont{Sackett}}, \bibinfo{journal}{Phys. Rev. A.}
  \textbf{\bibinfo{volume}{78}}, \bibinfo{pages}{023619}
  (\bibinfo{year}{2008}).

\bibitem[{\citenamefont{Segal et~al.}(2010)\citenamefont{Segal, Diot, Cornell,
  Zozulya, and Anderson}}]{Segal10}
\bibinfo{author}{\bibfnamefont{S.~R.} \bibnamefont{Segal}},
  \bibinfo{author}{\bibfnamefont{Q.}~\bibnamefont{Diot}},
  \bibinfo{author}{\bibfnamefont{E.~A.} \bibnamefont{Cornell}},
  \bibinfo{author}{\bibfnamefont{A.~A.} \bibnamefont{Zozulya}},
  \bibnamefont{and} \bibinfo{author}{\bibfnamefont{D.~Z.}
  \bibnamefont{Anderson}}, \bibinfo{journal}{Phys. Rev. A}
  \textbf{\bibinfo{volume}{81}}, \bibinfo{pages}{053601}
  (\bibinfo{year}{2010}).

\end{thebibliography}
\end{document}